\documentstyle[12pt,epsfig]{article}
\newcommand{\lbfig}[1]{\refstepcounter{fig} \label{#1} }
\newcounter{fig}

\newlength{\overeqskip}
\newlength{\undereqskip}
\newcommand{\nc}{\newcommand}
\nc{\be}{\begin{equation}}
\nc{\ee}{\end{equation}}
\nc{\bea}{\begin{eqnarray}}
\nc{\eea}{\end{eqnarray}}
\nc{\bi}[1]{\bibitem{#1}}
\nc{\lsim}{\mbox{\raisebox{-.6ex}{~$\stackrel{<}{\sim}$~}}}
\nc{\gsim}{\mbox{\raisebox{-.6ex}{~$\stackrel{>}{\sim}$~}}}
\nc{\nn}{\nonumber}
\def\Re{{\rm Re\,}}

\def\khx{\frac{k}{2H_{\rm I}}}

\begin{document}
%
%
\begin{titlepage}
\pagestyle{empty}
\baselineskip=16pt
\rightline{DAMTP-2000-44}
\rightline{CERN-TH/2000-200}
\rightline{IFIC/00-34}
\rightline{UNIL-IPT/00-11}
\rightline{July 2000}
\baselineskip=21pt
\begin{center}
     {\Large {\bf Primordial spectrum of gauge fields from inflation }}
\end{center}
\begin{center}
    Anne-Christine Davis,$^1$
    Konstantinos Dimopoulos,$^2$
     Tomislav Prokopec,$^3$\\ and\\
Ola T\"{o}rnkvist$^{1,4}$\\[1ex]
\baselineskip=16pt
$^1${\it Department of Applied Mathematics and Theoretical Physics,
University of Cambridge,}\\
{\it Wilberforce Road, Cambridge CB3~0WA, United Kingdom}\\

$^2${\it Instituto de F\'{\i}sica Corpuscular,
Universitat de Valencia/CSIC,}\\
{\it Apartado de Correos 2085, 46071 Valencia, Spain}\\

$^3${\it Universit\'e de Lausanne, Institut de Physique Th\'eorique,}\\
{\it BSP,  CH-1015 Lausanne, Switzerland}\\

$^4${\it CERN Theory Division, 1211 Gen\`{e}ve 23, Switzerland}
\end{center}

\baselineskip=21pt
\vskip 0.2 in

\centerline{ {\bf Abstract} }
\baselineskip=18pt
\vskip 0.5truecm
\noindent
We show that conformal invariance of gauge fields is naturally broken
in inflation, having as a consequence
amplification of gauge fields.
The resulting spectrum of the field strength is approximately
$B_{\ell}\propto {\ell}^{-1}$, where $\ell$ is the relevant coherence scale.
One realisation of our scenario is
scalar electrodynamics with
a scalar whose mass is large enough to evade observational constraints
-- the obvious candidates being supersymmetric partners
of the standard-model fermions.
Our mechanism also
leads naturally to amplification of the standard-model $Z$-boson field
due to its coupling to the electroweak
Higgs field. At preheating,
the spectrum of the $Z$ field is
transferred to the hypercharge field,
which remains frozen in the plasma
and is converted into a magnetic field at the electroweak phase transition.
With a reasonable model of field evolution one obtains a magnetic
field strength of the order of $10^{-29}$ Gauss on a scale of
100 pc, the size of
the largest turbulent eddy in a virialised galaxy.
Resonant amplification in preheating can lead to primordial
fields as large as $10^{-24}$ Gauss, consistent with the
seed field required for the galactic dynamo mechanism.
{}~\\[-3mm]
\rule{6cm}{.2mm}\newline
\noindent{$^1${\footnotesize acd@damtp.cam.ac.uk}, $^2${\footnotesize
kostas@flamenco.ific.uv.es}}\\*
\noindent{$^3${\footnotesize
Tomislav.Prokopec@ipt.unil.ch},
$^4${\footnotesize o.tornkvist@damtp.cam.ac.uk}}
\end{titlepage}

\baselineskip=20pt

%
%
\section{Introduction}

It is well known that during inflation the inflaton, being a light
scalar field, couples gravitationally and grows on superhorizon
scales. Likewise, (scalar) cosmological perturbations grow resulting
in a scale-invariant spectrum on superhorizon scales, providing one
of the most important predictions of inflation: seeds for formation of
large-scale structures in the Universe \cite{MukhanovChibisov,Hawking}.
The spectrum of gravitational waves generated during inflation has also
been studied in detail \cite{MukhanovChibisov} and is regarded as
perhaps the only signature of inflation that may be ``observed'' directly.
It is hence surprising
that the evolution of vector fields has not been considered with sufficient
care, especially in light of the exciting possibility that primordial
gauge fields from inflation could
produce the magnetic fields observed today in galaxies \cite{Kronberg}
and even
affect the cosmic microwave background radiation \cite{bfs}.

 It is usually assumed that gauge fields do not grow during inflation.
The reason is very simple: gauge fields do not couple gravitationally
to a conformally flat space-time. The metric of
conformally flat
space-times
can in general be written as $g_{\mu\nu}=a^2(\tau,\vec x)\eta_{\mu\nu}$,
where $\eta_{\mu\nu}={\rm diag}[1,-1,-1,-1]$ is the Minkowski metric.
The space-times of standard cosmology are all conformally flat.
Indeed, in inflationary de Sitter space-time the scale factor reads
$a(\tau)=-1/(H_{\rm I}\tau)$, where $H_{\rm I}$ is the Hubble parameter
in inflation; in the radiation
era $a\propto \tau$, while in the matter era  $a\propto \tau^2$.
This implies that, in order to get any amplification of gauge fields during
inflation, conformal invariance must be broken.

 Various proposals have been put forward in which conformal invariance of
gauge fields is broken. A particularly nice one is due to Dolgov \cite{dolgov},
who showed that the conformal anomaly of the gauge-field stress-energy tensor,
expressed by the triangle diagram, leads to particle production for gauge
fields even in a conformally flat space-time.
Turner and Widrow~\cite{tw} consider several ways of breaking conformal
invariance: (a) gravitational coupling of the photon, (b) anomalous coupling
of the photon to the axion, and (c)
coupling of the photon to a charged, massless
scalar field. In the former two cases, conformal anomaly is broken
by a non-standard coupling of the photon field. The case (c) has recently
been explored by Calzetta et al.\ and by Kandus et al.\ \cite{ckmw}.
These authors argue that, as a consequence of scalar charge separation
during inflation, charged domains form which proceed to source
electric currents in the radiation era,
leading to magnetic-field production. Their claim, that the magnetic
field thus created has sufficient strength to seed the galactic dynamo
mechanism \cite{dynamo}, has recently
been contested by Giovannini and Shaposhnikov \cite{gs}.

In this Letter we show that there is a {\it natural} way of
breaking conformal invariance of gauge fields which has not been considered
in the literature. Namely, the backreaction of a charged scalar field gives
the gauge field an effective mass, which breaks conformal invariance.

\section{Scalar electrodynamics}
\label{Scalar electrodynamics}

 We begin by considering scalar electrodynamics, which nicely illustrates
our mechanism of gauge-field amplification
from inflation. The idea can be
easily extended to include non-Abelian gauge fields and multi-component
scalar fields. The Lagrangian is
\begin{equation}
{\cal L}_{\rm \phi ED}=-\frac 14 g^{\mu\eta}g^{\nu\rho}F_{\mu\nu}F_{\eta\rho}
 + g^{\mu\nu}(D_\mu\phi)^\dagger D_\nu\phi - m_\phi^2\phi^\dagger\phi
  - \lambda_\phi(\phi^\dagger\phi)^2~,
\label{gf.1}
\end{equation}
where $D_\mu=\partial_\mu-ieA_\mu$ is the covariant derivative,
$F_{\mu\nu}=\partial_\mu A_\nu-\partial_\nu A_\mu$ is the gauge field strength,
$g^{\mu\nu}=a^{-2}\eta^{\mu\nu}$ is a conformally flat metric, and
$m_\phi$ and $\lambda_\phi$ are the mass and self-coupling of the scalar field
$\phi$, respectively. The experimental constraint on the $\phi$ mass,
$m_\phi\gg M_{\rm EW}\simeq 100$~GeV, is easily satisfied. The obvious
candidates for $\phi$ are supersymmetric partners of the standard-model
leptons and quarks.
In addition, we require that  $m_\phi\ll H_{\rm I}$, so that the charged
scalar field may grow during inflation.

 The relevant equation of motion for the photon field is obtained from
Eq.~(\ref{gf.1}) as the Lagrange equation
$\partial_\nu(\delta_{\partial_\nu A_\mu}(\sqrt{-D_g}{\cal L}_{\rm \phi ED}))
-\delta_{A_\mu}(\sqrt{-D_g}{\cal L}_{\rm \phi ED})=0$, where
$D_g={\rm det}[g_{\mu\nu}]=-a^8$. After some algebra
one finds that the mode equation for the transverse component of the photon
field in the Hartree approximation and in unitary gauge can be recast as
 \begin{equation}
\left(\partial_\tau^2 +\vec k^{\,2}
   + e^2a^2\langle\rho^2\rangle \right) {\cal A}_{\vec k} = 0~,
\label{gf.2}
\end{equation}
where $\rho^2 = 2\phi^\dagger \phi$,
$\vec k$ is the comoving momentum, $e$ electric charge
($e^2/4\pi\equiv \alpha\approx 1/137$), and $\tau$ the conformal time,
which is related to the ``physical'' time $t$ as $dt=a(\tau)d\tau$.
The average $\langle\cdot \rangle$ should be computed in the Hartree
approximation by subtracting the vacuum contribution,
so that in the vacuum $\langle\rho^2\rangle\rightarrow 0$
and Eq.~(\ref{gf.2}) reduces to
that of a harmonic oscillator. The solutions are then travelling harmonic
waves,
\begin{equation}
{\cal A}^{(\pm)}_{\rm vac}=\left(\frac{2\pi V}{k}\right)^{\frac{1}{2}}
   e^{\mp ik\tau}~,
\label{gf.3}
\end{equation}
where $k=|\vec{k}|$ and we used the (Wronskian) normalisation of the mode
functions, ${\bf W}[{\cal A}^{(+)}_{\rm vac},{\cal A}^{(-)}_{\rm vac}]$
$=4\pi iV$, appropriate for periodic boundary conditions in a box of volume
$V$. The modulus of the mode functions in
Eq.~(\ref{gf.3})
does not depend on time, which
is a consequence of conformal invariance being restored when
$\langle\rho^2\rangle\rightarrow\langle\rho^2\rangle_{\rm vac} = 0$.

For simplicity we shall consider only homogeneous space-times such that
the scale factor $a$ in Eq.~(\ref{gf.2}) may be taken to be
\begin{equation}
a=\left\{\begin{array}{lll}
 - 1/(H_{\rm I}\tau) & \mbox{for}~ \tau\leq -H_{\rm I}^{-1}
&\mbox{(inflation),}\\*
      H_{\rm I}\tau & \mbox{for}~ \tau\geq H_{\rm I}^{-1} &
\mbox{(radiation)}.\end{array}\right.
\label{gf.4}
\end{equation}
This
corresponds to the following smooth matching at the inflation-radiation
transition: $a(-H_{\rm I}^{-1})=a(H_{\rm I}^{-1})$ and
$da(-H_{\rm I}^{-1})/d\tau=da(H_{\rm I}^{-1})/d\tau$.

\subsection{Breakdown of conformal invariance}

The backreaction term in Eq.~(\ref{gf.2}) is generated during inflation
because of the gravitational coupling of the scalar field. In the Hartree
approximation, the backreaction is described by elastic scattering
processes only, which should be
a good approximation in inflation. We also assume that damping is negligible
in inflation, since most of the amplification is
on superhorizon scales beyond the reach of dissipative processes.

The Hartree term
$\langle\rho^2\rangle$ for a light,
charged scalar field can be estimated as follows. From the scalar equation of
motion, one infers that during inflation the scalar field grows
until its effective mass
essentially reaches the Hubble parameter. More precisely, we have
$m_\phi^2+3\lambda_\phi \langle \rho^2 \rangle\leq 2H_{\rm I}^2$,
so that
$e^2\langle\rho^2\rangle/H_{\rm I}^2\leq 2 e^2/(3\lambda_\phi)\simeq
1/(16\lambda_\phi)$. When the limit is saturated, the growth of $\phi$ modes
becomes exponentially suppressed. We assume that this occurs well before
the end of inflation.
The solutions of Eq.~(\ref{gf.2}) may then be conveniently
written in terms of Hankel functions as follows:
\begin{equation}
{\cal A}^{(i)}_{\vec k} = \pi\left(-\tau V\right)^\frac{1}{2}
   H^{(i)}_\nu(-k\tau)\,,\quad i=1,2,\qquad
 \nu^2=\frac{1}{4}-\frac{e^2\langle\rho^2\rangle}{H_{\rm I}^2}
 \quad {\rm (inflation)}.
\label{gf.5}
\end{equation}
For the
mode-function normalisation we impose the Wronskian condition
${\bf W}\left[{\cal A}^{(1)}_{\vec k},{\cal A}^{(2)}_{\vec k}\right]=4\pi iV$
so that, as $\tau\rightarrow -\infty$, the mode functions reduce to the vacuum
mode functions discussed above:
${\cal A}^{(1,2)}_{\vec k}\rightarrow {\cal A}_{\rm vac}^{(\pm)}$.
At later stages of inflation, however, the mode functions get ``squeezed''
on superhorizon scales as
\begin{equation}
{\cal A}_{\vec k}^{(j)}
\;\stackrel{k\vert \tau\vert \ll 1}{\longrightarrow}
\; i (-1)^j \Gamma(\nu)\left(\frac{2V}{k}\right)^\frac{1}{2}
 \left(-\frac{k\tau}{2}\right)^{\frac{1}{2}-\nu}
\,,\qquad   j=1,2 .
\label{gf.6}
\end{equation}
Since $\Re[\nu]<1/2$, the amplitudes ${\cal A}_{\vec k}^{(j)}$ actually
decrease, while the momenta $\partial_\tau{\cal A}_{\vec k}^{(j)}$ grow
(see Fig.~\ref{figure1}).
This evolution preserves the uncertainty relation for the photon
field operators, which can be equivalently expressed by the (conserved)
Wronskian.

After inflation the squeezed photon state evolves according to
complicated physics associated with the processes of
preheating and thermalisation of the Universe.
The details of the evolution are model dependent and
deserve further study \cite{ddpt},
especially since these processes have not been
fully considered in the context of gauge fields
(see, however, Ref.~\cite{BaackeHeitmannPatzoldBrusteinOakninFinelli}).
In the following section we shall make a crude approximation and
assume that the scalar field decays instantaneously at the end
of inflation, leading to sudden restoration of conformal symmetry.
The effects of preheating and thermalisation are considered in
section~\ref{preheating},
where we
show that, under certain conditions,
preheating results in additional amplification of the superhorizon modes.

\section{Photon field amplification from inflation}
\label{amplification in inflation}

Here we assume that soon after inflation the $\phi$ field decays
nonadiabatically such that $\langle\rho^2\rangle$ quickly approaches the
vacuum value
$\langle\rho^2\rangle_{\rm vac} = 0$,
rendering Eq.~(\ref{gf.2}) conformally invariant.
The solutions in the radiation era are then linear combinations of
the travelling waves ${\cal A}_{\rm vac}^{(\pm)}$ of Eq.~(\ref{gf.3}).
Particle production occurs since the effective mass parameter undergoes
a nonadiabatic change at the inflation-radiation transition. This change
can be nonadiabatic even when the
squared mass $e^2\langle\rho^2\rangle$
changes adiabatically. Indeed, at the beginning of radiation era the scale
factor evolves nonadiabatically, {\it i.e.}
$\vert (\partial_\tau a^2)/a^2\vert^2
= 4H_{\rm I}^2 \gg e^2\langle\rho^2\rangle$, so that the results
concerning the photon field amplification
in the simple
case under consideration are quite generic.

\begin{figure}[htp]
\begin{center}
\epsfig{file=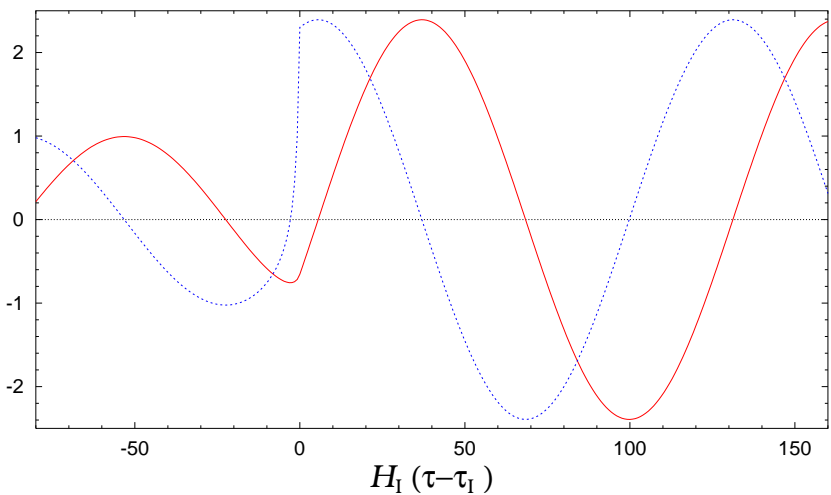,
width=4.5in
}
\\*
\end{center}
\vskip -0.1in
\lbfig{figure1}
\caption[fig1]{%
\small Evolution of (the imaginary part of) the
gauge-field amplitude ${\rm Im}\,[{\cal A}_{\vec{k}}]$ (red solid curve)
and its
momentum ${\rm Im}\,[\partial_\tau{\cal A}_{\vec{k}}]$ (blue dotted curve)
for $k/H_{\rm I}=0.05$ near the
inflation-radiation transition, which occurs at
$\tau=\tau_{\rm\scriptscriptstyle I}
=\pm H_{\rm I}^{-1}$. As a consequence of broken
conformal invariance, toward the end of inflation
the amplitude decreases and the derivative increases, leading
to an enhanced amplitude in the radiation era. The normalisation is
chosen so that both functions would have amplitude equal to unity
in the conformally
evolving case, and we use $\nu=0.2$ for the purpose of illustration.}
\end{figure}
Since the vacuum of inflation does not correspond to the
late-time vacuum of
the radiation era, the field content of Eq.~(\ref{gf.5}) may be
interpreted from the point of view of a late-time observer by a smooth
matching of the mode functions ${\cal A}_{\vec{k}}^{(i)}$ at the end of
inflation to a linear combination of travelling waves in the radiation
era,%
\begin{eqnarray}
{\cal A}^{(1)}_{\vec k} (- H_{\rm I}^{-1}) &=&
   \alpha_{\vec k}{\cal A}_{\vec k}^{(+)}(H_{\rm I}^{-1})
   + \beta_{-\vec k}^* {\cal A}_{-\vec k}^{(-)}(H_{\rm I}^{-1})
\nonumber\\
\partial_\tau{\cal A}^{(1)}_{\vec k}(-H_{\rm I}^{-1}) &=&
   \alpha_{\vec k}\partial_\tau{\cal A}_{\vec k}^{(+)}(H_{\rm I}^{-1})
   + \beta_{-\vec k}^* \partial_\tau{\cal A}_{-\vec k}^{(-)}(H_{\rm I}^{-1})~,
\label{gf.7}
\end{eqnarray}
and similarly for ${\cal A}^{(2)}_{\vec{k}}={\cal A}^{(1)\ast}_{\vec{k}}$.
Here $\alpha_{\vec k}$ and
$\beta_{\vec k}$ are the Bogoliubov coefficients that
relate the creation and annihilation operators in the radiation era
to those of (de Sitter) inflation.%
\footnote{The convention used here is consistent with the Bogoliubov
coefficients being defined through the following relations:
$a_{\vec{k}}=\alpha_{\vec{k}}\bar{a}_{\vec{k}}
+\beta_{\vec{k}}\bar{b}^{\dagger}_{-\vec{k}}$,
 $b_{\vec{k}}=\alpha_{\vec{k}}\bar{b}_{\vec{k}}
+\beta_{\vec{k}}\bar{a}^{\dagger}_{-\vec{k}}$, where barred operators
are for inflation and unbarred for radiation era.}
With ${\cal A}_{\vec k}^{(\pm)}={\cal A}_{\rm vac}^{(\pm)}$ the matching
problem is easily solved. On superhorizon scales we have
\begin{eqnarray}
\left(\begin{array}{c} \alpha_{\vec{k}}\\*\beta_{\vec{k}}
\end{array}\right)
&=&\mp e^{i{k}/{H_{\rm I}}}
\frac{\Gamma(\nu)}{4\sqrt{\pi}}
\left(\frac{1}{2}-\nu\right)\left(\khx\right)^{-\frac{1}{2}-\nu}
+{\cal O}(\left(k/2H_{\rm I}\right)^{-\frac{1}{2}+\nu})~,
\label{gf.8}
\end{eqnarray}
so that at late time in radiation the linearly independent
mode functions ${\cal A}_{\vec{k}}$ and ${{\cal A}_{\vec{k}}}^\ast$,
where ${\cal A}_{\vec{k}} \equiv \alpha_{\vec{k}}{\cal A}_{\vec{k}}^{(+)}
+ \beta_{-\vec{k}}^\ast {\cal A}_{-\vec{k}}^{(-)}$~, are proportional to
$k^{-1-\nu}$. When $\nu\approx 1/2$ the superhorizon modes
${\cal A}_{\vec{k}}\propto k^{-3/2}$ make up an almost scale-invariant
spectrum, similar to that of a massless scalar field.
The amplification by a factor
$\beta_{\vec{k}}\propto
(k/ H_{\rm I})^{-1/2-\nu}$ compared to the Minkowski vacuum
spectrum (see Fig.~\ref{figure1})
is a consequence of the large value of the momentum $\partial_\tau
{\cal A}_{\vec{k}}^{(j)}$ at the end of inflation, as can be inferred
from Eq.~(\ref{gf.6}).
This is the maximum amplification that can be attained at
the inflation-radiation transition, since sudden transition is the one
with maximal nonadiabaticity. However, it is not hard to show that
a softer matching,
for example onto $\langle\rho^2\rangle\sim H^2 =(H_{\rm I}\tau^{2})^{-2}$,
 leads to almost as strong an
amplification on superhorizon scales (for $\nu\approx 1/2$), namely
${\cal A}_{\vec{k}}\propto  k^{-\frac{1}{2}-2\nu}$. An obvious candidate for
such a scenario is ``gravitational'' preheating \cite{FordSpokoiny}.
We emphasise that, in a transition between asymptotic
''in'' and ''out'' states with the property $\langle\rho^2\rangle_{\rm in}
= \langle\rho^2\rangle_{\rm out}=0$, the expectation value
$\langle\rho^2\rangle$ will typically be non-zero during any period when
the quantum operator $\rho^2$ evolves non-adiabatically, leading to
amplification of gauge fields.

\section{Photon field amplification in preheating}
\label{preheating}

Here we present a simplified model of preheating in which the spectrum of
photons from inflation is further amplified on superhorizon scales, so long as
inelastic scattering and dissipation can be neglected. We assume that,
at late stages when thermalisation takes place, the $\phi$ field decays
nonadiabatically, imprinting the photon-field spectrum onto the plasma
as described in section~\ref{amplification in inflation}. How realistic
this assumption is will be the subject of a separate study~\cite{ddpt}.

Resonant amplification of the photon field may result when the scalar
field $\phi$ couples to the inflaton and grows resonantly such that the photon
mass term $e^2\langle\varphi^2\rangle$ ($\varphi^2=2a^2\phi^\dagger\phi$)
acquires an oscillatory component, leading to (secondary) resonant
amplification of the photon field $A$. This resonant amplification resembles
stochastic resonance~\cite{ProkopecRoosKofmanLindeStarobinsky} in which all
superhorizon modes are equally amplified. We shall now make a crude estimate
of the amplification factor.

 We assume massless chaotic inflation with a quartic interaction term
$\lambda_ss^4/4$, where $\lambda_s\simeq 10^{-13}$ as specified by the COBE
satellite microwave background radiation measurements. Furthermore,
the inflaton
$s$ couples to $\phi$ with a term $hs^2\phi^\dagger\phi$
such that $hs^2\leq 2H^2(s)$ during inflation, setting an upper limit on
$h$. It is now easy to show that the oscillating
inflaton $s$ decays into $\phi$ through a narrow resonance with
the quality factor $q_\phi\approx h s_0^2/4\omega_{\rm I}^2\leq 1/200$,
where $s_0 \approx 0.3M_{\rm P}$ is the inflaton amplitude at the end
of inflation and $\omega_{\rm I}
=c_n\sqrt{\lambda_s}\,s_0$ ($c_n\approx 0.847$)
is the inflaton frequency. Since at very early stages the photon-field quality
factor $q_A=e^2
\langle\varphi^2\rangle_{\rm osc}/4\omega_{\rm I}^2\ll q_\phi$,
where $\langle\varphi^2\rangle_{\rm osc}$ is the amplitude of the oscillatory
component of $\langle\varphi^2\rangle$, energy density $\rho$
is much more
efficiently transferred from $s$ to $\phi$ than it is from $\phi$ to $A$.
As time passes, $\langle\varphi^2\rangle_{\rm osc}$ grows, leading to more
efficient resonant production of the photon field $A$.
Quite generically, the photon-field superhorizon modes begin to grow when
$q_A\sim 1$ is attained~\cite{ProkopecRoosKofmanLindeStarobinsky}. At that
time $\tau_1$,
$\rho_{A}(\tau_1)\leq\rho_{\phi}(\tau_1)
\sim 12\lambda_\phi\omega_{\rm I}^4/e^4$.
The modes stop growing when
$\rho_\phi\sim\rho_A\sim \rho_s\simeq \omega_{\rm I}^4/(4\lambda_sc_n^4)$.
On the other hand, a resonating vector mode grows as
${\cal A}_{\vec k}\propto e^{\mu_A\omega_{\rm I}\tau}$ and hence
$\rho_A\propto e^{2\mu_A\omega_{\rm I}\tau}$,
implying the following estimate of
the amplification factor $\chi$ for superhorizon modes during preheating
\begin{equation}
\chi\sim
\left(\frac{\rho_s}{\rho_{A}(\tau_1)}\right)^\frac{1}{2}
\sim \frac{e^2}{4c_n^2} \frac{1}{\sqrt{3\lambda_\phi\lambda_s}}
\sim 10^5 .
\label{gf.9}
\end{equation}
We assume that the photon field at this point freezes in due to large
conductivity. The purpose of this simple analysis was to show that
superhorizon modes are amplified during preheating, a rough estimate
of the amplification factor being Eq.~(\ref{gf.9}). To obtain a more accurate
value, the problem should be reanalysed using numerical techniques.
We have also studied the evolution of the spectrum
by matching the inflation mode functions
to those in preheating, ${\cal A}_{\rm preh}^{(\pm)}
 =\left(2\pi V/\omega_{\rm I}\right)^{{1}/{2}}e^{(\mu_A\mp i)\omega_I\tau}$.
At the end of preheating, the
spectrum is ${\cal A}_{\rm preh}\propto k^{-\nu}$,
as
in Eq.~(\ref{gf.6}),
with an additional
amplification factor given by Eq.~(\ref{gf.9}). A final matching of
the preheating photon modes onto the massless photon modes in radiation
results in the spectrum ${\cal A}_{\rm rad}\propto k^{-1-\nu}$ computed in
section~\ref{amplification in inflation}, again with the additional
amplification factor~(\ref{gf.9}).

\section{Magnetic field production in the standard model}
\label{Magnetic field production in the standard model}

 In this section we show that the amplification mechanism,
which we illustrated with the example of
scalar electrodynamics, is operative for the standard-model
$Z$ field just as it is for any gauge field
that couples to a light scalar field. Namely,
in the $Z$-field
mode equation ({\it cf.} Eq.~(\ref{gf.2}))
 \begin{equation}
\left(\partial_\tau^2 +\vec k^{\,2}
   + \frac{m_Z^2}{v^2}a^2\langle\rho^2\rangle \right) {\cal Z}_{\vec k} = 0
\label{gf.10}
\end{equation}
conformal invariance is broken by the
Hartree
backreaction term of the
standard-model Higgs field $\Phi$.
Here $v=246$~GeV and
$\rho^2=2\Phi^\dagger\Phi$.
The restriction $3\lambda_H\langle\rho^2\rangle\leq 2H_{\rm I}^2$
implies that,
in order to obtain substantial amplification
 of superhorizon modes $(\nu\gsim 1/3)$,
we require
$\lambda_H\gsim 0.66$.
Equivalently, the Higgs-boson mass must satisfy the lower bound
$m_H= v\,\sqrt{2\lambda_H}\gsim
280$ GeV.
The spectrum reads
 \begin{equation}
{\cal Z}_{\vec{k}} \equiv \alpha_{\vec{k}}{\cal Z}_{\vec{k}}^{(+)}
  + \beta_{-\vec{k}}^\ast {\cal Z}_{-\vec{k}}^{(-)} \propto k^{-1-\nu}
\qquad {\rm with}\quad
  \nu^2=\frac{1}{4}-\frac{m_Z^2}{v^2}\frac{\langle\rho^2\rangle}{H_{\rm I}^2}~,
\label{gf.11}
\end{equation}
where ${\cal Z}^{(\pm)}_{\vec k}\equiv {\cal Z}^{(\pm)}_{\rm vac}
        =\left({2\pi V}/{k}\right)^{{1}/{2}} e^{\mp ik\tau}$
({\it cf.} Eq.~(\ref{gf.3})), and
$\alpha_{\vec{k}}$ and $\beta_{\vec{k}}$ are given in Eq.~(\ref{gf.8}).
The non-Abelian component of the $Z$ field, {\em i.e.\ }$W^{(3)}$,
develops a magnetic mass and becomes screened
as the Universe
thermalises. Only its Abelian component, the hypercharge field
$Y=-\sin\theta_{\rm W} Z$, survives and freezes into the plasma.
Here $\theta_{\rm W}$ is the Weinberg angle ($\sin\theta_{\rm W}
\approx 0.23$).
After the electroweak transition it is the photon field
which remains unscreened
with amplitude $A=\cos\theta_{\rm W} Y$.
Note that, although the $Z$ field and the photon field are orthogonal,
the resulting photon spectrum is identical to that of the original $Z$ field,
apart from an amplitude suppression factor $\sin\theta_{\rm W}
\cos\theta_{\rm W}\approx 0.42$.

\begin{figure}[tbp]
\begin{center}
\epsfig{file=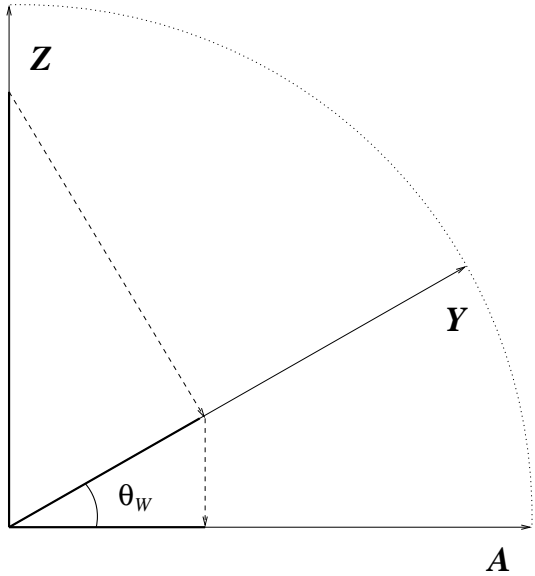, width=7.cm}
\\*
\end{center}
\lbfig{polarisers}
\caption[fig2]{%
\small Projection of the primordial spectrum of the standard-model $Z$ field
first onto the hypercharge field $Y$ (at the inflation-radiation transition)
and then onto the photon field $A$ (at the electroweak phase transition).}
\end{figure}
The situation is analogous to that of light polarisers
(see Fig.~\ref{polarisers}). With two
orthogonal light polarisers no light passes. When a third polariser
is inserted at an angle $\theta$ (with respect to the second polariser)
however, some of the photons do pass. The photon amplitude is reduced
by $\sin\theta\cos\theta$, just as in the Z-field case.
The main advantage of the amplification mechanism presented here
is its naturalness. Indeed, no fields are required
except those of the standard model and inflation.

One should also consider
the possibility
of additional enhancement of the field $Y=-\sin\theta_{\rm W} Z$
on superhorizon scales
during preheating. For this purpose,
we envision a hybrid-inflation model in which
inflation is driven by the vacuum energy of a GUT Higgs field.
After inflation ends, this field begins to oscillate, leading to
resonant production of some superheavy gauge fields $X_i$.
As a consequence
the hypercharge field $Y$, which couples to the $X_i$,
grows as well, and freezes in when the Universe thermalises.
The details of this amplification mechanism are model-specific and
will be presented elsewhere \cite{ddpt}.
Finally, we
emphasise that the superheavy gauge fields
grow resonantly during preheating and suddenly decay after preheating
when they become supermassive. These are required
properties of the scalar field in the scalar-electrodynamics model
considered in sections~\ref{Scalar electrodynamics}
and~\ref{amplification in inflation}.

\section{Magnetic field spectrum}
\label{Magnetic field spectrum}

We now consider the possibility that primordial gauge fields from
inflation can provide the necessary seed for the galactic dynamo
mechanism \cite{dynamo} and thereby explain the observed presence of
micro-Gauss-strength
magnetic fields in a large number
of spiral galaxies \cite{Kronberg}.

The dynamo mechanism amplifies a weak, coherent seed field using the
differential rotation of a galaxy in conjunction with the turbulent motion
of ionised gas.
The length scale relevant for
the operation of the galactic dynamo
is the size
of the largest turbulent eddy, $\sim 100$ pc.
On this scale, there is a minimum strength
of the magnetic seed field
for which the dynamo can operate. Estimates
of this minimal strength lie in the range $10^{-23}$--$10^{-19}$
G for a universe with critical matter density and zero cosmological
constant. The lower bound
can be relaxed \cite{dlt} to about $10^{-30}$ G for a flat, low-density
universe with a dark-energy component (e.g. a cosmological
constant or quintessence),
which
appears to
be favoured by recent
results from supernova observations and balloon experiments
\cite{supernova,boomerang}.

The dynamo scale ($100$ pc) corresponds to a comoving scale today
of $\ell_{c}\sim 10$~kpc $\approx 1.56\times 10^{36}$ GeV$^{-1}$
before gravitational collapse and galaxy formation \cite{dlt}.
Because the concentration
of matter into a galaxy brings about an amplification of
magnetic fields by a factor
$(\rho_{\rm gal}/\rho_0)^{2/3}
\approx 5\times10^3$, the bounds that should be imposed
on the scale $\ell_{c}$ are $B_{\rm seed}\gsim 2\times 10^{-27}$ G
for a universe with critical
matter density and $B_{\rm seed}\gsim 2\times 10^{-34}$ G
for a flat, dark-energy dominated, low-density universe.

We consider first the spectrum resulting from scalar electrodynamics
(section~\ref{amplification in inflation}) and the $Z$-field case
(section~\ref{Magnetic field production in the standard model})
and neglect for the time being further amplification which may result
from parametric resonance during preheating.
At the end of inflation,
the scale factor corresponds to an equivalent temperature $T$ given by
$H_{\rm I} = \pi  g_\ast^{1/2}(T) T^2 / ({\sqrt{90}M_{\rm P}})$,
where
$g_\ast(T)\sim 10^2$ is the number of relativistic degrees of freedom.
Assuming first that
the primordial magnetic field is frozen in the plasma from
the time of its creation,
the relevant length scale for the magnetic field at the end of inflation
is $\ell=\ell_{\rm c} T_0/T$, where $T_0\approx2.73$ K is the temperature
today.

For $\bar{k}\equiv 2\pi/\ell\ll H_{\rm I}$,
the Bogoliubov coefficients $\alpha_{\vec{k}}$ and $\beta_{\vec{k}}$ given
by Eq.~(\ref{gf.8}) are maximal for
$\nu=\bar{\nu}\approx 1/2-
1/\ln (2H_{\rm I}/\bar{k})$.
For this optimal value of $\nu$ one obtains the mode function
\begin{equation}
|{\cal A}_{\vec{k}}|
\sim \frac{1}{4}\Gamma(\bar{\nu})(\frac{1}{2}-\bar{\nu})
\left(\frac{V}{H_{\rm I}}\right)^\frac{1}{2}
 \left(\frac{k}{2H_{\rm I}}\right)^{-\bar{\nu}-1}.
\label{gf.13}
\end{equation}
The magnetic field correlated on a particular scale $\ell$
is defined by
\begin{eqnarray}
&&B_\ell^2 =
\langle B_i(\ell,\vec{x}) B_i(\ell,\vec{x})\rangle -
\langle B_i(\ell,\vec{x}) B_i(\ell,\vec{x})\rangle_{\rm vac}~,\\*
&&B_i(\ell,\vec{x}) =\frac{3}{4\pi\ell^3}
\int_{|\vec{y}-\vec{x}|\leq\ell} d^3y B_i(\vec{y})~,
\label{gf.14}
\end{eqnarray}
where the average $\langle\cdot\rangle$ is
taken over Fock space as well as the position $\vec{x}$.
For $\ell H_{\rm I}\gg 1$ one finds
\begin{equation}
B_\ell=3\times 2^{\bar{\nu}-2}
\frac{\Gamma(\bar{\nu})}{\pi}
(\frac{1}{2}-\bar{\nu})
H_{\rm I}^2 (\ell H_{\rm I})^{\bar{\nu}-\frac{3}{2}}~.
\label{gf.15}
\end{equation}

The magnetic field then is assumed to be frozen in the plasma
such that today $B_{\ell_{\rm c}} = B_{\ell}\,
(T_0/T)^2$.
Since roughly $B_{\ell_{\rm c}}\propto T$,
both $B_{\ell_{\rm c}}$ and $B_{\ell}$ are larger for
higher inflation scale $H_{\rm I}\propto T^2$.
Taking $H_{\rm I}\sim10^{13}$ GeV (corresponding to
$T\sim 10^{15}$ GeV) we find
$B_{\ell}\sim10^{22}$ G at the end of inflation,
and $B_{\ell_{\rm c}}\sim 10^{-34}$ G today at a comoving scale of 10 kpc
(see Fig.~\ref{spectra}).

\begin{figure}[htbp]
\begin{center}
\epsfig{file=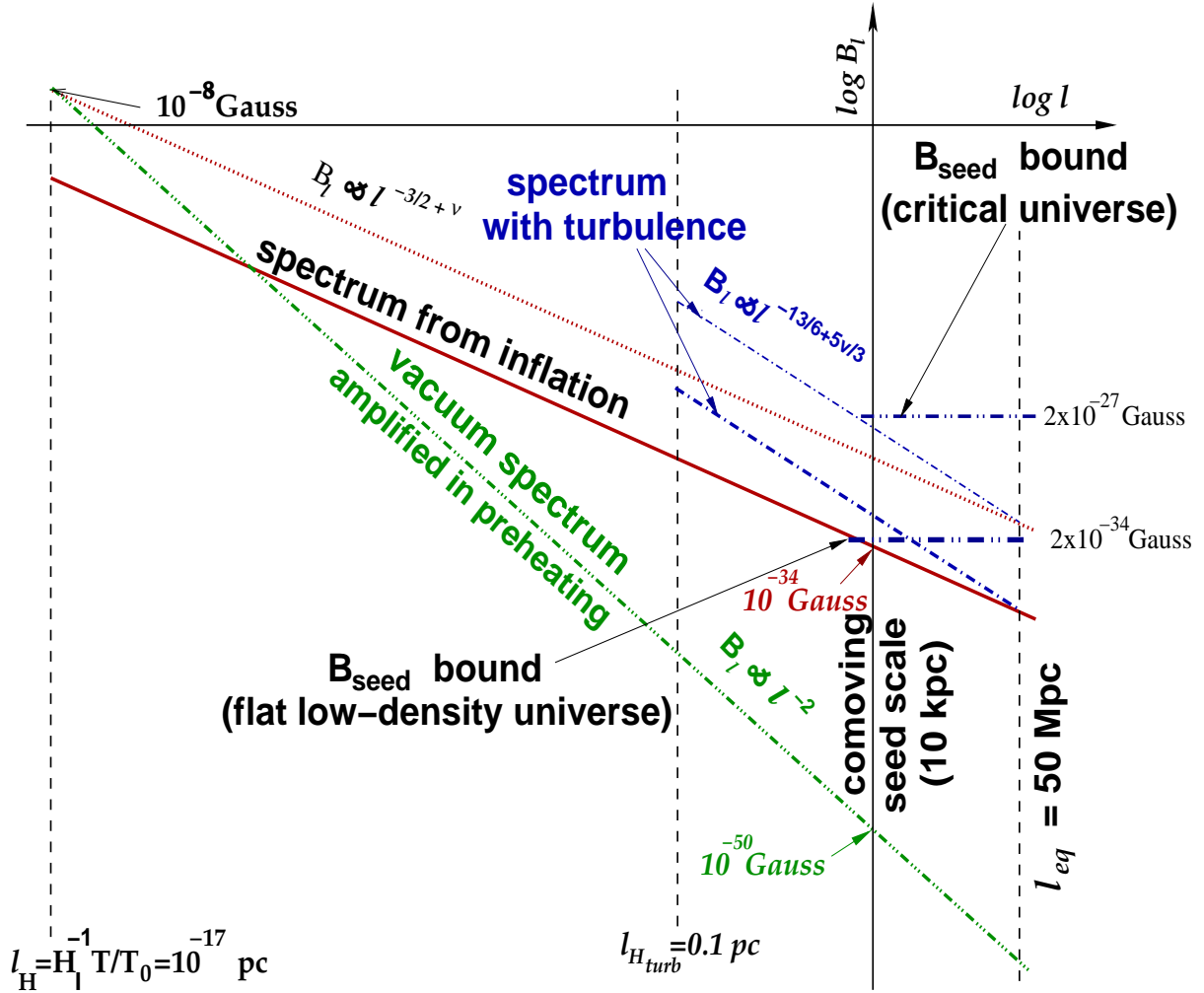, height=5.5in,width=6.5in}\\*
\end{center}
\vskip -0.3in
\lbfig{spectra}
\caption[fig3]{%
\small Magnetic-field spectra and relevant seed-field bounds. In green
({\em dash-dot-dot-dot}) we show the vacuum spectrum
$B_\ell\propto \ell^{-2}$ obtained from preheating, assuming an amplification
factor of $10^5$. At the comoving scale
$\ell_c\sim 10$~kpc, $B_{\ell_c}\sim 10^{-50}$~G.
In red ({\em dots and solid}) the spectrum $B_\ell\propto \ell^{-3/2+\nu}$
from inflation in our mechanism is shown, {\em with} and
{\em without} preheating amplification.
For this spectrum, $B_{\ell_c}\sim 10^{-29}$~G
and $B_{\ell_c}\sim 10^{-34}$~G, respectively. We also show
({\em blue dash-dots}) the spectrum enhanced by helical turbulence
(at $\ell_c\sim 10$~kpc an enhancement of about 20 is obtained).
This is to be compared with
the dynamo bounds rescaled by a factor $5\times 10^3$ (see main text)
$B_{\rm seed} \gsim 2\times 10^{-27}$~G for a universe with critical
matter density, and $B_{\rm seed} \gsim 2\times 10^{-34}$~G for a flat,
low-density universe.}
\end{figure}
In section~\ref{preheating} we showed that this field strength can be further
amplified by a factor of about $10^5$ through parametric resonance with an
oscillating scalar field, leading to $B_{\ell_{\rm c}}\sim 10^{-29}$~G.
A supplementary increase in field strength is obtained if we assume that
the magnetic field does not freeze into the plasma upon creation, but
rather that its correlation length grows quicker than the scale factor,
as is the case for helical turbulence \cite{Son}. Such a causal mechanism can
only operate on a given comoving scale after this scale has reentered the
horizon. One can show that the growth of correlations
due to turbulent evolution leads to an additional amplification of about
$(\ell_{\rm eq}/\ell_{\rm c})^{2(1-\nu)/3}$,
where $\ell_{\rm eq}\sim 50$~Mpc denotes the equal matter-radiation
horizon today. For $\ell_{\rm c}=10$~kpc the amplification is about 20.
For other types of turbulence the amplification may be somewhat
smaller.

In Fig.~\ref{spectra} we display the spectrum $B_\ell$
of the magnetic field for the different
histories of amplification discussed in this section and compare with
the comoving $\ell_{\rm c}=10$ kpc seed-field bounds, which are
$B_{\rm seed}\gsim 2\times 10^{-27}$~G in a
universe with critical matter density (presently disfavoured by observations)
and $B_{\rm seed}\gsim 2\times 10^{-34}$~G in a flat, low-density universe
dominated by dark energy. The corresponding field strengths and bounds
in a newborn galaxy can be obtained by multiplying with a factor
$5\times 10^3$.

To conclude, in this Letter we have presented a generic
mechanism for production of gauge fields during
inflation which predicts an almost scale-invariant spectrum for
gauge fields, comparable with the scale-invariant spectrum of
cosmological perturbations. When applied to electromagnetism, the resulting
magnetic-field spectrum $B_\ell\propto\ell^{\,\nu -3/2}\sim \ell^{-1}$
can provide strong enough
seed fields to explain the origin of galactic magnetic fields.

\section*{Acknowledgements}

TP wishes to thank M.\ Giovannini and M.\ Shaposhnikov  for
useful discussions.
Support for
KD was provided by DGICYT grant PB98-0693 and by the
European Union under contract ERBFMRX-CT96-0090; for OT
by the European Union under contract ERBFMBI-CT97-2697 and by
a CERN visiting fellowship. We gratefully acknowledge travel support
by the U.K. PPARC.

%
%

\nc{\ap}[3]    {{\it Ann.\ Phys.\ }{{\bf #1}, {#2} {(#3)}}}
\nc{\ibid}[3]  {{\it ibid.\ }{{\bf #1}, {#2} {(#3)}}}
\nc{\jmp}[3]   {{\it J.\ Math.\ Phys.\ }{{\bf #1}, {#2} {(#3)}}}
\nc{\np}[3]    {{\it Nucl.\ Phys.\ }{{\bf #1}, {#2} {(#3)}}}
\nc{\pl}[3]    {{\it Phys.\ Lett.\ }{{\bf #1}, {#2} {(#3)}}}
\nc{\pr}[3]    {{\it Phys.\ Rev.\ }{{\bf #1}, {#2} {(#3)}}}
\nc{\prd}[3]    {{\it Phys.\ Rev.\ }D {{\bf #1}, {#2} {(#3)}}}
\nc{\prep}[3]  {{\it Phys.\ Rep.\ }{{\bf #1}, {#2} {#3)}}}
\nc{\prl}[3]   {{\it Phys.\ Rev.\ Lett.\ }{{\bf #1}, {#2} {(#3)}}}
\nc{\spjetp}[3]{{\it Sov.\ Phys.\ JETP }{{\bf #1}, {#2} {(#3)}}}
\nc{\zetp}[3]  {{\it Zh.\ Eksp.\ Teor.\ Fiz.\ }{{\bf #1}, {#2} {(#3)}}}

\end{document}